\documentclass[twocolumn]{autart}    

\usepackage{epsfig}
\usepackage{graphicx}
\usepackage{subfig}

\usepackage[utf8]{inputenc}
\usepackage{amsmath}
\usepackage{amssymb}
\usepackage{amsfonts}
\usepackage{siunitx}
\usepackage{psfrag}
\usepackage{enumitem}
\usepackage{xcolor}
\usepackage{notoccite}
\usepackage{threeparttable}
\usepackage{verbatim}
\usepackage{epstopdf}
\usepackage{todonotes}

\usepackage{natbib}

\usepackage{stfloats}

\newtheorem{Remark}{Remark}
\newtheorem{Theorem}{Theorem}
\newtheorem{Definition}{Definition}
\newtheorem{Lemma}{Lemma}
\newtheorem{Corollary}{Corollary}

\newcommand{\abs}[1]{\left\vert #1 \right\vert}
\newcommand{\norm}[1]{\left\Vert #1 \right\Vert}
\newcommand{\R}{{\mathbb R}}  

\newcommand{\bb}[1]{{\color{black} #1}}

\usepackage{hyperref}

\begin{document}

\begin{frontmatter}

\title{On $\mathcal{L}_{\infty}$ string stability of nonlinear bidirectional asymmetric heterogeneous platoon systems} 


\author[Rome]{Julien Monteil}  
\author[Baiae]{Giovanni Russo}  \ead{giovanni.russo1@ucd.ie} 
\author[Baiae]{Robert Shorten}  
\address[Rome]{IBM Research - Ireland Lab. Address: IBM Technology Campus, Mulhuddart, Dublin, 15.}             
\address[Baiae]{University College Dublin, School of Electrical Engineering. Address: Belfield, Dublin, $4$.}        


\begin{abstract}
This paper is concerned with the study of bidirectionally coupled platoon systems. The case considered is when the vehicles are heterogeneous and the coupling can be \bb{nonlinear and asymmetric}. For such systems, a sufficient condition for  $\mathcal{L}_{\infty}$ string stability is presented. The effectiveness of our approach is illustrated via a numerical example, where it is shown how our result can be recast as an optimization problem, \bb{allowing to design the control protocol for each vehicle independently on the other vehicles and hence} leading to a bottom-up approach for the design of string stable systems \textcolor{black}{able to track a time-varying reference speed.}
\end{abstract}

\end{frontmatter}
\section{Introduction}

Platoon systems designate a class of network systems where automated vehicles, typically arranged in a string, cooperate via some distributed control protocol, or coupling, in order to travel along the longitudinal direction~\citep{Lev_Ath_66}. The vehicles need to attain a configuration where a common driving speed is achieved and, at the same time, some desired vehicle-to-vehicle distance is kept. Typically, the distributed protocols needs to be designed so as to ensure string stability of the platoon system, see e.g.~\citep{1341587,5406127,5208241,5680944,7299636}. Intuitively, if the system is string stable, then: (i) vehicles can attain and keep the desired configuration; (ii) the effects of disturbances are attenuated along the string.

A platoon system is said to be {\em unidirectional} (also termed as leader-follower topology) if the control protocol on each vehicle only takes as input information coming from the vehicles ahead, while it is said to be {\em bidirectional} if the control protocol takes as input information coming from the vehicles ahead and behind, see e.g.~\citep{5406127,nieuwenhuijze2010string}. Recently, see e.g. \citep{hao2012robustness,martinec2016necessity,HERMAN201713,herman2017scaling}, {\em asymmetric} bidirectional control algorithms have been considered, where the information from the vehicles ahead might be weighted differently than the information from the vehicles behind. Also, the platoon system is said to be {\em homogeneous} if the vehicles are all identical, {\em heterogeneous} otherwise.
\subsection*{Literature review} 
Historically, work on string stability can be traced back to~\citep{peppard1974string} and to the {\em California PATH} program, see e.g.~\citep{sheikholeslam1990longitudinal}. A convenient way to formalize the concept of string stability is via the use of $p$-signal norms. \bb{The concept of $\mathcal{L}_p$ string stability has been originally introduced in~\citep{swaroop1996string}, where a number of sufficient conditions ensuring this property were also given. In such a paper, $\mathcal{L}_p$ string stability was defined for interconnected systems with no external disturbance. Recently, a similar formalism has been used in~\citep{knorn2014passivity}, where $\mathcal{L}_p$ string stability has been defined for systems affected by external disturbances. Another convenient way to formalize $\mathcal{L}_p$ string stability has been introduced in~\citep{6515636}. In such a paper, the definition of string stability is given for systems where the first vehicle is affected by an external disturbance imposed by the leading vehicle. Essentially, following~\citep{6515636}, the platoon system with the first vehicle affected by the disturbance is $\mathcal{L}_p$ string stable if the $\mathcal{L}_p$ signal norm of the local error vector between the current and target states of the system is upper bounded by certain class $\mathcal{K}$ functions.} 

\bb{The notions of $\mathcal{L}_2$ and $\mathcal{L}_\infty$ string stability are particularly useful for applications. As noted in \citep{6515636}, the use of $\mathcal{L}_2$ string stability is motivated by requirements of energy dissipation along the system, while the notion of $\mathcal{L}_\infty$ string stability is related to the maximum vehicle overshoot \citep{STUDLI20172511}. This concept, in turn, has a direct interpretation in terms of vehicle collisions. For linear systems, studying $\mathcal{L}_2$ string stability, while lacking the interpretation in terms of collision avoidance, is analytically convenient as results can be stated in terms of the $\mathcal{H}_{\infty}$ system norm of the transfer function.

In \citep{5406127} it is shown how $\mathcal{L}_2$ string stability can be achieved for a linear platoon by allowing inter-vehicle communications and in~\citep{ploeg2014controller}, the design of a $\mathcal{L}_{2}$ string stable cooperative adaptive cruise controller, making use of a feed-forward term, is presented for linear systems where the disturbance is on the first vehicle. In the linear setting, in \citep{nieuwenhuijze2010string} a $\mathcal{L}_2$ string stability definition in the $z$-domain is given for homogeneous platoon systems  and this is used to analyze the performance of bidirectional constant time headway control policies. In particular, one of the main findings is that the use of a bidirectional structure can result in a better disturbance attenuation when compared to a predecessor-follower strategy. Recently, in~\citep{948631,7963246}, it has been shown that constant time-headway policies can be used to enhance $\mathcal{L}_{2}$ string stability in linear platoon systems. Also, in~\citep{hao2012robustness}, the robustness to external disturbances is investigated for linear, heterogeneous, platoon systems where vehicles are modeled as double integrators and where the disturbance is a sinusoidal function. In particular, quantitative comparisons between unidirectional, bidirectional and asymmetric bidirectional control protocols are presented in the paper and it is shown how asymmetric bidirectional control protocols can have a beneficial effect on string stability. Indeed, one of the main findings of this paper is that asymmetric weights on the velocity feedback enhances robustness of the platoon system. The implications of asymmetric bidirectional control protocols on disturbance scaling and $\mathcal{L}_{2}$ string stability have been further investigated for linear platoons in~\citep{HERMAN201713}, and in~\citep{martinec2016necessity} via a wave-based control approach. In~\citep{728529}, nonlinear spacing policies are introduced for automated heavy-duty vehicles and string stability is proven on the linearized system. Instead, an approach to the design of nonlinear protocols for platoon systems has been presented in~\citep{knorn2014passivity}, where energy-based arguments are used to prove $\mathcal{L}_2$ string stability. This approach has been also expanded in~\citep{knorn2015scalability} to mitigate the effects of time-varying measurement errors on the platoon. Finally, in~\citep{7937859}, nonlinear control protocols are studied but only stability is considered rather than string stability, while consensus-based approaches are explored in~\citep{6891349} and \citep{7972982}, where exponential stability is considered in the case where some of the vehicles in the platoon are subject to speed restrictions. 

The literature on $\mathcal{L}_{\infty}$ string stability of linear platoon systems is sparse when compared to the literature on $\mathcal{L}_{2}$ string stability.  Conditions for $\mathcal{L}_{\infty}$ string stability of linear, unidirectional, platoon systems have been originally investigated in~\citep[Chapter $5$]{swaroop1996string}. Other works on $\mathcal{L}_{\infty}$ string stability of linear platoon systems include~\citep{Hedrick_99,801173,4610029,8318387}. In particular, in \citep{Hedrick_99,801173} unidirectional platoons with no external disturbances  are considered, while in \citep{4610029} the platoon does not have a leading vehicle and the use of ring interconnection topologies are explored, when only the first vehicle is affected by an external disturbance. Finally, in the recent work \citep{7879221}, the problem of studying $\mathcal{L}_{\infty}$ string stability for nonlinear homogeneous, unidirectional, platoons is investigated in the spatial domain and the methodology is illustrated by designing distributed protocols requiring each vehicle to use position, speed and acceleration from the leading vehicle.}
\subsection*{Contribution of this paper} 
In the context of the above literature, this paper offers the following contributions: \bb{(i) a novel sufficient condition for $\mathcal{L}_\infty$ string stability is presented for heterogeneous platoon systems coupled with nonlinear, asymmetric bidirectional control protocols and subject to disturbances. The string stability definition used in this paper generalizes a number of definitions commonly used in the literature (see Definition \ref{def:stability} and Remark \ref{rem:def}); (ii) The control policies devised following our theoretical results allow the platoon system to track a desired (possibly, non-constant) reference speed: this is particularly appealing for applications, where the reference speed might be used to e.g. set speed restrictions; (iii) It is shown how our theoretical results can be effectively used to design protocols guaranteeing $\mathcal{L}_\infty$ string stability of the platoon system. Namely, we show how the results can be recast as an optimization problem that allows to design the control protocol for each vehicle independently on the other vehicles.} 

\section{Notation and problem formulation}\label{Sec:2}

\subsection{Notation}
Let $v$ be an arbitrary $m$-dimensional vector, $A$ be a $m\times m$ matrix and $\Theta$ be a non-singular $m\times m$ matrix. By $\abs{v}_p$ we denote an arbitrary $p$-vector norm on $\R^m$, while $\norm{A}_p$ and $\mu_p(A)$ denote the matrix norm and matrix measure of $A$ induced by $\abs{\cdot}_p$, see e.g.~\citep{Vid_93} and Appendix~\ref{math_tools}. Then, $\abs{v}_{\Theta,p} = \abs{\Theta v}_p$ is also a vector norm and its induced matrix measure is equal to $\mu_{\Theta,p}(A) = \mu_p\left(\Theta A \Theta^{-1}\right)$. We also denote by $\sigma_{\max}(A)$ ($\sigma_{\min}(A)$) the largest (smallest) singular value of $A$, \textcolor{black}{by $[A]_s$ the symmetric part of $A$ and by $I_m$ the identity matrix of dimension $m$. The notation $A\succeq 0$ indicates that the matrix $A$ is positive semi-definite and $A \prec B$ that the matrix $A-B$ is negative definite.} Consider the signal $d(\cdot):\R_+\rightarrow \mathcal{D}\subseteq \R^{n}$. Then, the {\em supremum norm} of $d(\cdot)$ is denoted by $\norm{d(\cdot)}_{\mathcal{L}_\infty} = \sup_{t\ge 0}\abs{d(t)}_2$. A continuous function, $a(\cdot): \R_+ \rightarrow \R_+$ is: (i) a class-$\mathcal{K}$ function if $a(0) = 0$ and $a(\cdot)$ is strictly increasing; (ii) a class-$\mathcal{L}$ function if it monotonically decreases to $0$ as its argument tends to $+\infty$. A continuous function, $b(\cdot, \cdot): \R_{+} \times \R_{+} \rightarrow \R_{+}$, is  a class-$\mathcal{KL}$ function if $b(\cdot, t)$ is a class-$\mathcal{K}$ function $\forall t \ge 0$ and $b(\xi,\cdot)$ is a class-$\mathcal{L}$ function, $\forall\xi \ge 0$.

\subsection{System description}\label{sec:sys_descr}

We consider platoon systems of  $N > 1$ heterogeneous vehicles arranged along a string and following a leading vehicle (vehicle $0$). The dynamics of the $i$-th vehicle within the platoon is governed by:
\begin{equation}\label{eqn:generalext}
\dot x_i = f_i(x_i)  + \tilde u_i + d_i, \ \ \ x_i(t_0) = x_{i,0}, \ \ \ t_0 \ge 0,
\end{equation}
$i=1,\dots,N$. We consider the case where: (i) $x_i\in \R^n$; (ii) $f_i(x_i):\R^n\rightarrow \R^n$ are smooth functions; (iii) $\tilde u_i(t)$ is the distributed control protocol having the form
\begin{equation}\label{eqn:automated_asymmetric_general}
\begin{split}
\tilde u_i(t) &:= \tilde h_{i,i-1}(t,x_{i-1},x_i) + \varepsilon_i \tilde h_{i,i+1}(t,x_{i+1},x_i) \\
& + \textcolor{black}{\tilde{h}_i^{(0)}}(t,x_{i},x_0).
\end{split}
\end{equation}
In (\ref{eqn:automated_asymmetric_general}) the functions $\tilde h_{ij}:\R_+ \times \R^n \times \R^n \rightarrow \R^n$ are smooth coupling functions and $0 \le \varepsilon_i \le 1$ is the coupling gain between vehicle $i$ and the vehicle behind, i.e. vehicle $i+1$,~\citep{HERMAN201713}.  We say that (\ref{eqn:automated_asymmetric_general}) is: (i) an asymmetric control protocol, if $0<\varepsilon_i<1$; (ii) a predecessor-follower protocol, if $\varepsilon_i = 0$; (iii) a bidirectional protocol, if $\varepsilon_i = 1$. In (\ref{eqn:automated_asymmetric_general}), $x_0$ is the input received  from vehicle $0$ and the smooth function $\textcolor{black}{\tilde{h}_i^{(0)}}(\cdot,\cdot)$ represents a direct coupling, which requires a communication infrastructure, from the leading vehicle to vehicle $i$. If there is no communication between those vehicles, then $\textcolor{black}{\tilde{h}_i^{(0)}}(\cdot,\cdot) = 0$ by definition. The dynamics (\ref{eqn:generalext}) - (\ref{eqn:automated_asymmetric_general}) models a network of nonlinear heterogeneous vehicles coupled via time-dependent coupling functions. Explicitly including nonlinearities in the design of the control protocols is useful in certain applications such as platooning of heavy-duty vehicles where the nonlinearities at the vehicles cannot be neglected, see e.g. \citep{7286902} and references therein. Finally, $d_i(t)$ is an $n$-dimensional disturbance acting on the $i$-th vehicle (in the context of this paper a disturbance is an $n$-dimensional signal with all of its components being  piece-wise continuous), $d(t):=[d_1(t)^T,\ldots,d_N(t)^T]^T$ and $X(t):=[x_1(t)^T,\ldots,x_N(t)^T]^T$.

\subsection{Control goals}\label{sec:control_goals}

In order to introduce our results we define the unperturbed dynamics of (\ref{eqn:generalext}) - (\ref{eqn:automated_asymmetric_general}) as
\begin{equation}\label{eqn:unperturbedext}
\begin{split}
\dot y_i &= f_i(y_i)  + \tilde h_{i,i-1}(t,y_{i-1},y_i) \\
& + \varepsilon_i \tilde h_{i,i+1}(t,y_{i+1},y_i) + \textcolor{black}{\tilde{h}_i^{(0)}}(t,y_{i},x_0),
\end{split}
\end{equation}
and we denote by $Y(t):= [y_1(t)^T,\ldots,y_N(t)^T]^T$ the stack of all $y_i$'s. Also, the desired solution for (\ref{eqn:unperturbedext}) is denoted by $Y^{\ast}(t) := [y_1^{\ast}(t)^T,\ldots,\break y_N^{\ast}(t)^T]^T$, with $\dot{y}^{\ast}_{i} = f_i(y^{\ast}_{i})$, $\forall i=1,\ldots,N$. That is, $Y^{\ast}(t)$ corresponds to a desired configuration of the platoon system when there are no disturbances.  Our goal in this paper is to design the control protocols $\tilde u_i(\cdot)$ in (\ref{eqn:generalext}) so as to guarantee \textcolor{black}{disturbance} $\mathcal{L}_{\infty}$ string stability of the platoon system. This is formalized via the following definition, see also \citep{7879221}

\begin{Definition}\label{def:stability}
Consider the platoon system (\ref{eqn:generalext}) and assume that $Y^{\ast}(t)$ is a solution of its unperturbed dynamics (\ref{eqn:unperturbedext}). Then, (\ref{eqn:generalext}) is said to be \textcolor{black}{disturbance} $\mathcal{L}_{\infty}$ string stable if there exists a class $\mathcal{KL}$ function, $\alpha$, a class-$\mathcal{K}$ function, $\beta$, such that, for any disturbance $d(t)$ and initial conditions, we have, $\forall t \ge t_0$, 
$\textcolor{black}{{\sup_{i}\abs{x_i(t)- y_{i}^{\ast}(t)}}_{\textcolor{black}{2}}}\leq \alpha\left(\textcolor{black}{\sup_i\abs{x_i(t_0) - y_{i}^{\ast}(t_0)}_2},t\right)+\beta\left(\textcolor{black}{\sup_i}\norm{d_{\textcolor{black}{i}}(\cdot)}_{\mathcal{L}_{\infty}}\right)$.
\end{Definition}
\bb{\begin{Remark}\label{rem:def}
In the above definition, $\textcolor{black}{{\sup_{i}\abs{x_i(t)- y_{i}^{\ast}(t)}}_{\textcolor{black}{2}}}$ is upper bounded by the same functions $\alpha(\cdot,\cdot)$ and $\beta(\cdot)$ for any platoon length, $N$. That is, the bounds of the definition are independent on the number of vehicles. Definition \ref{def:stability} is stated via the input-to-state stability formalism, see e.g.~\citep{sontag2008input,Kha_02}, and disturbances are explicitly considered (in the definition given in e.g.~\citep{swaroop1996string} disturbances are not considered). Also, Definition \ref{def:stability} generalizes the definition given in~\citep{ploeg2014controller} as it allows to consider platoon systems with disturbances acting on any vehicle within the system.
\end{Remark}} 
\textcolor{black}{\begin{Remark}
Definition \ref{def:stability} can be equivalently stated in terms of $\abs{\cdot}_\infty$ by  noticing that:
\begin{equation*}
\begin{split}
\sup_{i}\abs{x_i(t)- y_{i}^{\ast}(t)}_{2} & \ge \sup_{i}\abs{x_i(t)- y_{i}^{\ast}(t)}_{\textcolor{black}{\infty}}\\
\sup_i\abs{x_i(0) - y_{i}^{\ast}(0)}_2 &\le \sqrt{n}\sup_i\abs{x_i(0) - y_{i}^{\ast}(0)}_\infty\\
\norm{d_i(\cdot)}_{\mathcal{L}_\infty} &\le \sqrt{n} \sup_i\sup_t\abs{d_i(t)}_\infty,
\end{split}
\end{equation*}
thus giving a bound that is  still independent on the number of vehicles, $N$.
\end{Remark}}
In what follows, systems fulfilling Definition \ref{def:stability} are simply termed as $\mathcal{L}_{\infty}$ string stable. \textcolor{black}{Finally,  we now establish a link between Definition \ref{def:stability} and the string stability definition given in \citep{swaroop1996string}.
\begin{Lemma}
Assume that the platoon system (\ref{eqn:generalext}) is $\mathcal{L}_{\infty}$ string stable in accordance to Definition \ref{def:stability}. Then, if $d_i(t) = 0$, $\forall t\ge0$ and $\forall i =1,\ldots,N$,  Definition \ref{def:stability} is equivalent to the definition given in~\citep{swaroop1996string}.
\end{Lemma}
\noindent {\em Proof.} The proof can be  obtained from} \citep[page $175$]{Kha_02} and it is omitted here for brevity. \qed

\section{Results}\label{sec:results}

With the result below, a sufficient condition for $\mathcal{L}_{\infty}$ \textcolor{black}{string} stability of the platoon system is given.

\begin{Theorem}\label{thm:predecessor_follower_general}
Consider the platoon system (\ref{eqn:generalext}) controlled by the distributed control strategy (\ref{eqn:automated_asymmetric_general}). Assume that the coupling functions $\tilde h_{i,i-1}$, $ \tilde h_{i,i+1}$, $\textcolor{black}{\tilde{h}_i^{(0)}}$ and the control gains $\varepsilon_i$ are designed in a way such that, $\forall i =1,\ldots,N$ and $\forall t\ge t_0$:
\begin{enumerate}
\item[{\bf C1 - }] $\tilde h_{i,i-1}\left(t,y^{\ast}_{i-1},y^{\ast}_{i}\right) =\tilde  h_{i,i+1}\left(t,y^{\ast}_{i+1},y^{\ast}_{i}\right) = 0$, $\textcolor{black}{h_i^{(0)}}\left(t,y^{\ast}_{i},x_{0}\right) = 0$;
\item[{\bf C2 - }] \textcolor{black}{for some $c \ne 0$, $\bar J >0$
\begin{equation}
\begin{split}
&\mu_{2}\left(\frac{\partial f}{\partial x_i}+\frac{\partial \textcolor{black}{\tilde{h}_i^{(0)}}}{\partial x_i}  +\frac{\partial \tilde h_{i,i-1}}{\partial x_i}+\varepsilon_i\frac{\partial \tilde h_{i,i+1}}{\partial x_i} \right) \le -c^2,\\
&\max\left\{\norm{\frac{\partial \tilde h_{i,i-1}}{\partial x_{i-1}}}_2,\norm{\frac{\partial \tilde h_{i,i+1}}{\partial x_{i+1}}}_2  \right\} \le \bar J,
\end{split}
\end{equation}}
$\forall x_i,x_{i-1},x_{i+1} \in\R^n$;
\item[{\bf C3 - }] $\varepsilon_i < \frac{c^2}{\bar J} -1$.
\end{enumerate}
Then, 
\textcolor{black}{\begin{equation*}
\begin{split}
 \sup_i\abs{x_i(t) -y_i^{\ast}(t)}_{\textcolor{black}{2}} 
& \le  e^{-\bar c^2(t-t_0)}\sup_i\abs{x_i(t_0)- y_i^{\ast}(t_0)}_{\textcolor{black}{2}} \\
&  + \frac{1- e^{-\bar c^2(t-t_0)}}{\bar c^2}\textcolor{black}{\sup_i\norm{d_i(\cdot)}_{\mathcal{L}_\infty}},
\end{split}
\end{equation*}}
$\forall t \ge t_0$, where $\bar c^2 := c^2 -\bar J(1+\max_i\varepsilon_i)$.
\end{Theorem}
\textit{Proof.}
See Appendix~\ref{sec:proof}. \qed

We now focus on the special case where: (i) the vehicles within the platoon in (\ref{eqn:generalext}) are modeled via a second order linear system with a disturbance acting on the acceleration; (ii) the desired platoon configuration is the configuration where  vehicles keep a desired distance from the vehicle ahead, while following a reference speed. In doing so, we denote by $\textcolor{black}{q_i}(t)$ and $v_i(t)$  the position and speed of the $i$-th vehicle ($x_i(t):= [\textcolor{black}{q_i}(t),v_i(t)]^T$) having as initial conditions $\textcolor{black}{q_i}(0)$ and $v_i(0)$. The \bb{(possibly) time-varying} reference speed is $v_0\bb{(t)}$ and, without loss of generality we assume that $q_0(0) =0$. Then, the position of the leading vehicle at time $t$ is denoted by $q_0(t)$ and we let $x_0(t):= [q_0(t),v_0\bb{(t)}]^T$ be the input from the leading vehicle to the platoon system. The dynamics of the $i$-th vehicle, $i=1,\ldots,N$, is governed by: 
 \begin{equation*}
 \begin{split}
& \dot{\textcolor{black}{q_i}} = v_i,\\
 & m_i\dot v_i = \bar u_i + \bar d_i,
 \end{split}
 \end{equation*}
 where $m_i$ is the mass of the $i$-th vehicle in the system, $\bar d_i(t)$ is a one-dimensional time-dependent disturbance on the vehicle  and $\bar u_i(t)$ is the decentralized control protocol for the $i$-th vehicle. In compact form we have:
\begin{equation}\label{eqn:general}
\begin{split}
\dot x_i &= F x_i  + u_i + d_i,
\end{split}
\end{equation}
$i = 1,\ldots,N$ and where: (i) $d_i(t):=1/m_i[0,\bar d_i(t)]^T$; (ii) $F=\begin{bmatrix}0&1\\0&0\end{bmatrix}$; (iii) $u_i(t) = [0, \bar u_i(t)]^T$. For notational convenience, we let
\begin{equation}\label{eqn:automated}
\begin{split}
\frac{1}{m_i}\bar u_i(t) :=& h_{i,i-1}(x_{i-1},x_i) + \varepsilon_i h_{i,i+1}(x_{i+1},x_i)\\
& + \textcolor{black}{h_i^{(0)}}(x_{i},x_0),\ \forall i=1,\ldots N.
\end{split}
\end{equation}
We define the  desired inter-vehicle distance between vehicle $i$ and the predecessor as $\delta_{i,i-1} > 0$ and we denote by $Y_d = [y_{d,1}^T,\ldots,y_{d,N}^T]^T$, $y_{d,i} := [q_0(t) - \delta_{i,0},v_0\bb{(t)}]^T$, the desired platoon configuration, where $\delta_{i,0} := \sum_{j=0}^{i-1}\delta_{j+1,j}$.  It is useful to introduce the positive constant $\alpha_i$ and make use of the matrices defined at the bottom of the page in (\ref{eqn:big_Jacobian}), where the dependency on the state variables has been omitted. We set by definition $J_{i,j}(\alpha_i,x_i,x_{j}) = 0$ whenever $i,j \notin\left\{1,\ldots,N\right\}$. Also, for all $i=1,\ldots,N$, it is convenient to define the matrix
\begin{equation}\label{eqn:T_matrix}
T_i := \left[\begin{array}{*{20}c}
1 & \alpha_i\\
0 & 1\\
\end{array}\right].
\end{equation}
Given this set-up, we can state the following.

\begin{figure*}[b]
\hrulefill
\begin{equation}\label{eqn:big_Jacobian}
\begin{array}{*{20}l}
J_{i,i}(\alpha_i,x_i,x_{i-1},x_{i+1},\varepsilon_i) := 
 \left[\begin{array}{*{20}c}
\alpha_i \frac{\partial \left(h_{i,i-1}+\varepsilon_i h_{i,i+1} + \textcolor{black}{h_i^{(0)}}\right)}{\partial \textcolor{black}{q_i}} & 1 - \alpha_i^2\frac{\partial \left(h_{i,i-1}+\varepsilon_i h_{i,i+1} + \textcolor{black}{h_i^{(0)}}\right)}{\partial \textcolor{black}{q_i}} + \alpha_i\frac{\partial \left(h_{i,i-1}+\varepsilon_i h_{i,i+1} + \textcolor{black}{h_i^{(0)}}\right)}{\partial v_i} \\
\frac{\partial \left(h_{i,i-1}+\varepsilon_i h_{i,i+1} + \textcolor{black}{h_i^{(0)}}\right)}{\partial \textcolor{black}{q_i}} & - \alpha_i\frac{\partial \left(h_{i,i-1}+\varepsilon_i h_{i,i+1} + \textcolor{black}{h_i^{(0)}}\right)}{\partial \textcolor{black}{q_i}} + \frac{\partial \left(h_{i,i-1}+\varepsilon_i h_{i,i+1} + \textcolor{black}{h_i^{(0)}}\right)}{\partial v_i}
\end{array}\right]\\
\\
J_{i,i-1}(\alpha_i,x_i,x_{i-1}) := \left[\begin{array}{*{20}c}
\alpha_i \frac{\partial h_{i,i-1}}{\partial \textcolor{black}{q_{i-1}}} & - \alpha_i^2\frac{\partial h_{i,i-1}}{\partial \textcolor{black}{q_{i-1}}} + \alpha_i\frac{\partial h_{i,i-1}}{\partial v_{i-1}} \\
\frac{\partial h_{i,i-1}}{\partial \textcolor{black}{q_{i-1}}} & - \alpha_i\frac{\partial h_{i,i-1}}{\partial \textcolor{black}{q_{i-1}}}+ \frac{\partial h_{i,i-1}}{\partial v_{i-1}}
\end{array}\right], \ \ 
J_{i,i+1}(\alpha_i,x_i,x_{i+1}) :=
 \left[\begin{array}{*{20}c}
\alpha_i \frac{\partial h_{i,i+1}}{\partial \textcolor{black}{q_{i+1}}} & - \alpha_i^2\frac{\partial h_{i,i+1}}{\partial \textcolor{black}{q_{i+1}}} + \alpha_i\frac{\partial h_{i,i+1}}{\partial v_{i+1}} \\
\frac{\partial h_{i,i+1}}{\partial \textcolor{black}{q_{i+1}}} & - \alpha_i\frac{\partial h_{i,i+1}}{\partial \textcolor{black}{q_{i+1}}}+ \frac{\partial h_{i,i+1}}{\partial v_{i+1}}
\end{array}\right].\\
\end{array}
\end{equation}
\end{figure*}

\begin{Corollary}\label{thm:string_automatic_vehicles}
Consider the platoon system (\ref{eqn:general}) controlled by the control strategy (\ref{eqn:automated}). Assume that the coupling functions $h_{i,i-1}$, $h_{i,i+1}$, $\textcolor{black}{h_i^{(0)}}$ and the control gains $\varepsilon_i$ are designed in a way such that, $\forall i =1,\ldots,N$:
\begin{enumerate}
\item[{\bf C1 - }] $h_{i,i-1}\left(y_{d,i-1},y_{d,i}\right) = h_{i,i+1}\left(y_{d,i+1},y_{d,i}\right) = 0$,  $\textcolor{black}{h_i^{(0)}}(y_{d,i},x_{0}) = 0$;
\item[{\bf C2 - }]  \textcolor{black}{for some $\alpha_i >0$, $c \ne 0$, $\bar J >0$ 
\begin{equation}
\begin{split}
&\mu_{2}\left(J_{i,i}(\alpha_i,x_i,x_{i-1},\varepsilon_i)\right) \le -c^2,\\
&\max\left\{\norm{J_{i,i-1}(\alpha_i,x_i,x_{i-1})}_{2}, \norm{J_{i,i+1}(\alpha_i,x_i,x_{i+1})}_{2}\right\} \le \bar J,
\end{split}
\end{equation}}
$\forall x_i,x_{i-1},x_{i+1} \in\R^2$;
\item[{\bf C3 - }] $\varepsilon_i < \frac{c^2}{\bar J} -1$.
\end{enumerate}
Then the system is $\mathcal{L}_{\infty}$ string stable. Moreover, $\forall i=1,\ldots,N$: 
\textcolor{black}{\begin{equation}\label{eqn:upper_bound}
\begin{split}
 \sup_i\abs{x_i(t) - y_{d,i}(t)}_{2} & \le K e^{-\bar{c}^2t}\sup_i\abs{x_i(0)-y_{d,i}(0)}_2 \\
&+ K \frac{1-e^{-\bar{c}^2t}}{\bar{c}^2} \sup_i\norm{d_i(\cdot)}_{\mathcal{L}_\infty},
\end{split}
\end{equation}}
$\forall t \ge 0$ and where $\bar c^2 := c^2 -\bar J(1+\max_i\varepsilon_i)$ and $K := \frac{\max_i\left\{\sigma_{\max}(T_i)\right\}}{\min_i\left\{\sigma_{\min}(T_i)\right\}}$.
\end{Corollary}
\textit{Proof.}
See Appendix~\ref{sec:proof}. \qed

\begin{Remark}
\textcolor{black}{{\bf C1} of Theorem \ref{thm:predecessor_follower_general}  implies that the desired platoon configuration is a solution of (\ref{eqn:generalext}). For platoon dynamics (\ref{eqn:general}), as the desired inter-vehicle distances are independent on the reference speed, {\bf C1} can be fulfilled via a constant inter-vehicle spacing policy. In turn, as noted in e.g. \citep{herman2017scaling,STUDLI20172511}, for systems with more than one integrator in the open loop, string stability cannot be guaranteed by these policies. This is typically overcome by allowing vehicles to take as input information from the leader. {\bf C2} implies that, in order to achieve string stability, the distributed protocols need to be designed so as to minimize the matrix measures/norms of the matrices in (\ref{eqn:big_Jacobian}). {\bf C3} states that the asymmetric coupling gains need to be designed as a function of the bounds obtained from {\bf C2}.}  
\end{Remark}
\textcolor{black}{\begin{Remark}
In Section \ref{sec:3}, we show how  the fulfillment of {\bf C2} and {\bf C3} can be recast as an optimization problem that allows to design the control protocol for each vehicle independently on the other vehicles. In turn, this leads to a {\em bottom-up} approach in the design of the platoon system.
\end{Remark}
Finally, we now define the matrix
$
H_{i,i}(\alpha_i,x_i,x_{i+1}) :=
 \left[
 \begin{array}{*{20}c}
\alpha_i \frac{h_{i,i+1}}{\partial \textcolor{black}{q_i}} & - \alpha_i^2\frac{\partial h_{i,i+1}}{\partial \textcolor{black}{q_i}} + \alpha_i\frac{\partial h_{i,i+1}}{\partial v_i} \\
\frac{\partial h_{i,i+1}}{\partial \textcolor{black}{q_i}} & - \alpha_i\frac{\partial h_{i,i+1}}{\partial \textcolor{black}{q_i}} + \frac{\partial h_{i,i+1}}{\partial v_i}
\end{array}
\right]
$
and, omitting the dependence on state variables for notational convenience, let $J_{i,i} = A_{i,i} + \varepsilon_i H_{i,i}$ (with the matrix $A_{i,i}$ defined accordingly). 
\begin{Corollary}\label{cor:string_stability_predecessor_follower}
Assume that, for the platoon system (\ref{eqn:general}) - (\ref{eqn:automated}): (i)  conditions {\bf C1}, {\bf C2} and {\bf C3} of Corollary \ref{thm:string_automatic_vehicles} are fulfilled for some $\alpha_i$, $c$, $\bar J$ and with $0 < \varepsilon_i \le 1$; (ii) the coupling functions $h_{i,i+1}$ are designed so that, for some $c_h \ne 0$, $\mu_2(H_{i,i}) \le - c_h^2$. Then, the corresponding predecessor-follower strategy obtained by setting $\varepsilon_i = 0$ also ensures string stability of the platoon system.
\end{Corollary}
\textit{Proof.} Indeed, note that: (i) {\bf C1} is independent on $\varepsilon_i$; (ii) if {\bf C3} is fulfilled for some $0 <\varepsilon_i \le 1$, then it is also satisfied when $\varepsilon_i$ is set to $0$. Thus, we only need to show that, if $\mu_{2}\left(J_{i,i}(\alpha_i,x_i,x_{i-1},\varepsilon_i)\right) \le -c^2$, then also $\mu_{2}\left(J_{i,i}(\alpha_i,x_i,x_{i-1},0)\right) \le -c^2$. In order to do so, note that, for any $0 \le \varepsilon_i' < \varepsilon_i$ it hold that: $\mu_2\left({A_{i,i} + \varepsilon_i' H_{i,i}}\right) =  \mu_2\left({A_{i,i} + \varepsilon_i H_{i,i} + (\varepsilon_i' - \varepsilon_i )H_{i,i}}\right) \le \mu_2\left({A_{i,i} + \varepsilon_i H_{i,i}}\right) + (\varepsilon_i - \varepsilon_i')\mu_2\left( H_{i,i}\right) \le - c^2 -(\varepsilon_i - \varepsilon_i') c_h^2$,
thus proving the result.}
\qed

\section{Numerical Validation}\label{sec:3}

We now use Corollary \ref{thm:string_automatic_vehicles} to design distributed control strategies ensuring string stability of the platoon system (\ref{eqn:general}). In order to do so, we consider the protocol (\ref{eqn:automated}) with: 
\begin{equation}\label{eqn:protocolNL2}
\begin{split}
h_{i,i-1} &= g_i\left(q_{i-1} - q_i - \delta_{i,i-1}\right) + K_{vi}\left(v_{i-1}-v_i\right),\\
 h_{i,i+1} &= g_i\left(q_{i+1}-q_i + \delta_{i+1,i}\right)+K_{vi}\left(v_{i+1}-v_{i}\right),\\
\textcolor{black}{h_{i}^{(0)}} &= K_{pi,0}(q_0 - q_i -\delta_{i,0}) + K_{vi,0}\left(v_{0}-v_i\right),
\end{split}
\end{equation}
and
\begin{equation}\label{eqn:protocolNL2_fcns}
g_i\left(x\right):=K_{pi,1}\tanh\left(K_{pi,2}x\right).
\end{equation} 
In (\ref{eqn:protocolNL2}) - (\ref{eqn:protocolNL2_fcns}), the parameters $K_{vi}$, $K_{pi,0}$, $K_{vi,0}$, $K_{pi,1}$ and $K_{pi,2}$ are control gains that will be tuned by applying Corollary \ref{thm:string_automatic_vehicles}. In the protocol, the nonlinear functions for the position coupling between vehicles (i.e. the functions $g_i(\cdot)$'s) are inspired by the optimal velocity model in~\citep{bando1995dynamical}, which mimics the human acceleration profile in a car-following configuration and embeds comfort considerations. Also, as in e.g.~\citep{1341587,6891349,herman2017scaling,5208241} we make use of a direct coupling between the leading vehicle $0$ and the $i$-th vehicle in the platoon. \bb{The key difference between (\ref{eqn:protocolNL2}) - (\ref{eqn:protocolNL2_fcns}) with respect to such papers is that the coupling functions $g_i$'s are nonlinear and our results are global results for string stability.}

In order to apply Corollary~\ref{thm:string_automatic_vehicles}, we first note that \textbf{C1} is verified by construction for the protocol~(\ref{eqn:protocolNL2}) - (\ref{eqn:protocolNL2_fcns}) and that $ \frac{\partial g_i(x)}{ \partial x}=K_{pi,1} K_{pi,2}\left(1-\tanh^2\left(K_{pi,2}x\right)\right)$ ${0\le \frac{\partial g_i(x)}{ \partial x} \le K_{pi,1}K_{pi,2}:= \bar g_i}$). Also, in this case, the matrices $J_{i,i}$, $J_{i,i+1}$ and $J_{i,i-1}$ are given at the bottom of the next page in (\ref{eqn:big_Jacobian_example2}). We recast the problem of finding a set of control gains fulfilling \textbf{C2} and \textbf{C3} for (\ref{eqn:big_Jacobian_example2}) as the optimization problem (\ref{optimisation}) of Appendix~\ref{sec:opt}. \textcolor{black}{Such a problem was solved via the Matlab {\em CVX} module, using the {\em Sedumi} solver. In particular, by setting $\varepsilon_i =1$ the following set of parameters satisfying the conditions of Corollary \ref{thm:string_automatic_vehicles} was found: $K_{pi,0}=0.50$, $K_{vi}=0.15$, $K_{vi,0}=0.38$, \textcolor{black}{$K_{pi,1}=0.50$, $K_{pi,2}=0.35$}. \textcolor{black}{The {\em CVX} code used to solve the optimization problem (\ref{optimisation}) of Appendix~\ref{sec:opt} is available online at \url{https://github.com/julien-monteil/automatica}.} Also, by means of Corollary \ref{cor:string_stability_predecessor_follower}, we know that the predecessor-follower strategy obtained by simply changing $\varepsilon_i$ to $0$ also guarantees $\mathcal{L}_{\infty}$ string stability of the platoon system.}
\begin{figure*}[b]
\hrulefill
\begin{equation}\label{eqn:big_Jacobian_example2}
\begin{array}{*{20}l}
J_{i,i}(\alpha_i,x_i,x_{i-1},x_{i+1},\varepsilon_i) =
 \left[\begin{array}{*{20}c}
-\alpha_i \left((1+\varepsilon_i)\frac{\partial  g_i}{\partial \textcolor{black}{q_i}}+K_{pi,0}\right)& 1+\alpha_i^2\left((1+\varepsilon_i)\frac{\partial  g_i}{\partial \textcolor{black}{q_i}}+K_{pi,0}\right)-\alpha_i \left(K_{vi}(1+\varepsilon_i)+K_{vi,0}\right) \\
-(1+\varepsilon_i)\frac{\partial  g_i}{\partial \textcolor{black}{q_i}}-K_{pi,0} & \alpha_i \left((1+\varepsilon_i)\frac{\partial  g_i}{\partial \textcolor{black}{q_i}}+K_{pi,0}\right) - K_{vi}(1+\varepsilon_i)-K_{vi,0}
\end{array}\right],\\
\\
J_{i,i-1}(\alpha_i,x_i,x_{i-1}) =
 \left[\begin{array}{*{20}c}
\alpha_i \frac{\partial  g_i}{\partial \textcolor{black}{q_{i-1}}} & - \alpha_i^2\frac{\partial  g_i}{\partial \textcolor{black}{q_{i-1}}}+\alpha_i K_{vi}\\
\frac{\partial g_i}{\partial \textcolor{black}{q_{i-1}}} & - \alpha_i \frac{\partial  g_i}{\partial\textcolor{black}{q_{i-1}}}+K_{vi}
\end{array}\right], \ \ \ \  J_{i,i+1}(\alpha_i,x_i,x_{i+1}) =  \left[\begin{array}{*{20}c}
\alpha_i \frac{\partial  g_i}{\partial \textcolor{black}{q_{i+1}}} & - \alpha_i^2\frac{\partial  g_i}{\partial \textcolor{black}{q_{i+1}}}+\alpha_i K_{vi}\\
\frac{\partial g_i}{\partial \textcolor{black}{q_{i+1}}} & - \alpha_i \frac{\partial  g_i}{\partial\textcolor{black}{q_{i+1}}}+K_{vi}
\end{array}\right].\end{array}
\end{equation}
\end{figure*}
 \bb{Simulations, illustrated in Figure~\ref{fig_1}, were performed by means of Matlab, using the second order Euler numerical method. In the simulations, we set $v_0=20~\si{\metre\per\second}$ and $\delta_{i,i-1}= \delta_{i+1,i} = 10~\si{\metre}$ (note that any other inter-vehicle distance and reference speed profile could be selected as the optimization problem in Appendix \ref{sec:opt} is independent on such parameters). In Figure~\ref{fig_1}, the time behavior is shown for the position and speed perturbations of a string of $N = 1000$ vehicles when \textcolor{black}{the perturbations $\bar{d_i}(t)=\eta_i 5\sin(t)\exp(-0.02t)$} are applied at time $t=0$ to $500$ randomly selected vehicles (the parameters $\eta_i\in[-1,1]$ are random scaling factors). This choice of $d_i(t)$'s physically corresponds to realistic but strong disturbances \citep{monteil2016robust}. Figure \ref{fig_1} clearly shows that, both the bidirectional and the predecessor-follower protocols designed so as to fulfill the conditions of Corollary \ref{thm:string_automatic_vehicles} and Corollary \ref{cor:string_stability_predecessor_follower}, ensure a $\mathcal{L}_{\infty}$ string stable behavior. Also, in accordance to e.g.~\citep{hao2012robustness,herman2017scaling,nieuwenhuijze2010string}, we found in the simulations that the bidirectional control exhibits a better disturbance rejection for both the position and speed deviations (that is, the peak of the position deviation is observed to be at \textcolor{black}{2.2~\si{\metre} for $\varepsilon_i =0$ and at 1.9~\si{\metre}} for $\varepsilon_i =1$, and the peak of the speed deviation is observed to be at \textcolor{black}{1.9~\si{\metre\per\second} for $\varepsilon_i =0$ and at 1.7~\si{\metre\per\second}} for $\varepsilon_i =1$).
\begin{figure}[thbp]
\centering
\begin{tabular}{cc} 
{\includegraphics[width=0.23\textwidth]{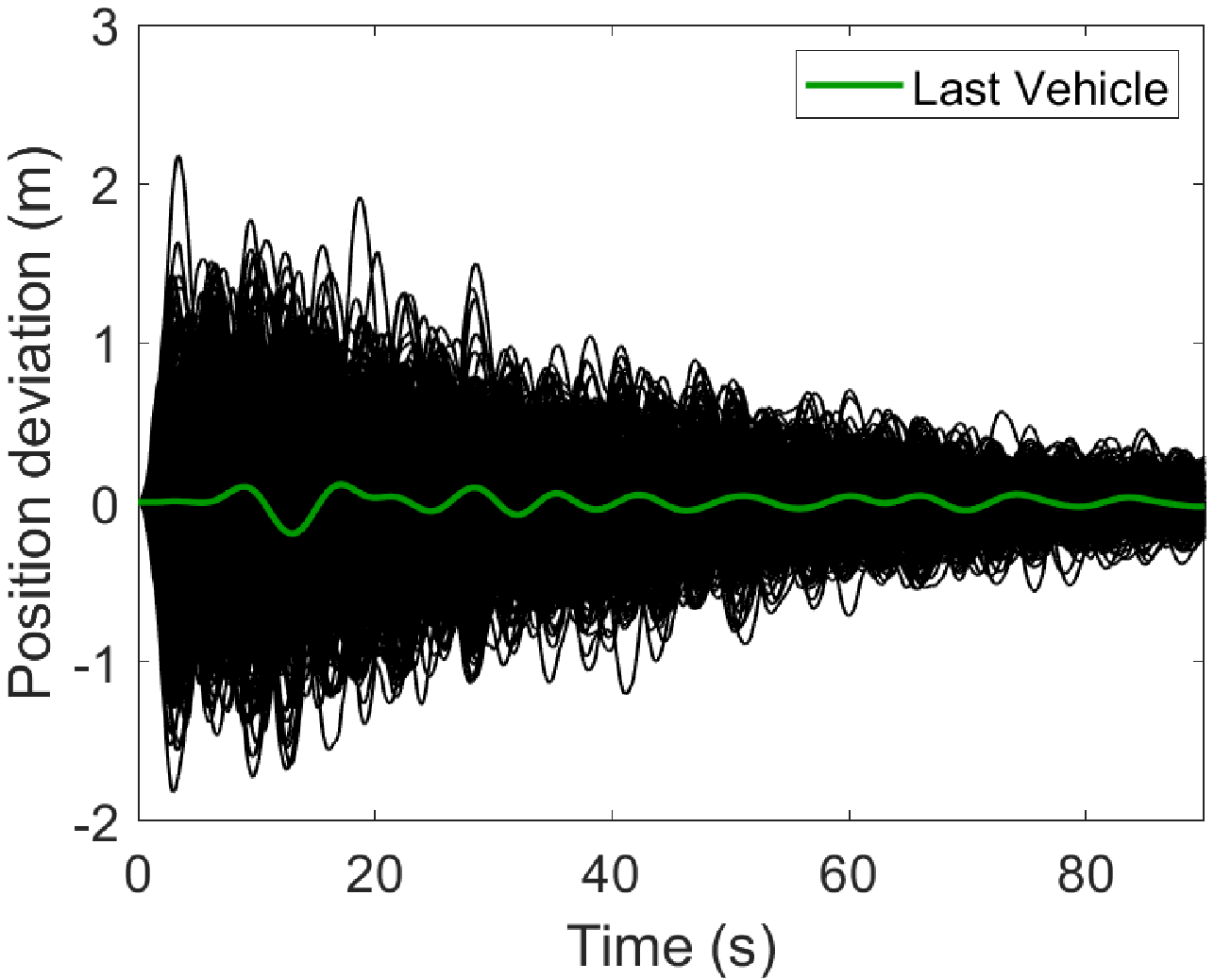}}&
{\includegraphics[width=0.23\textwidth]{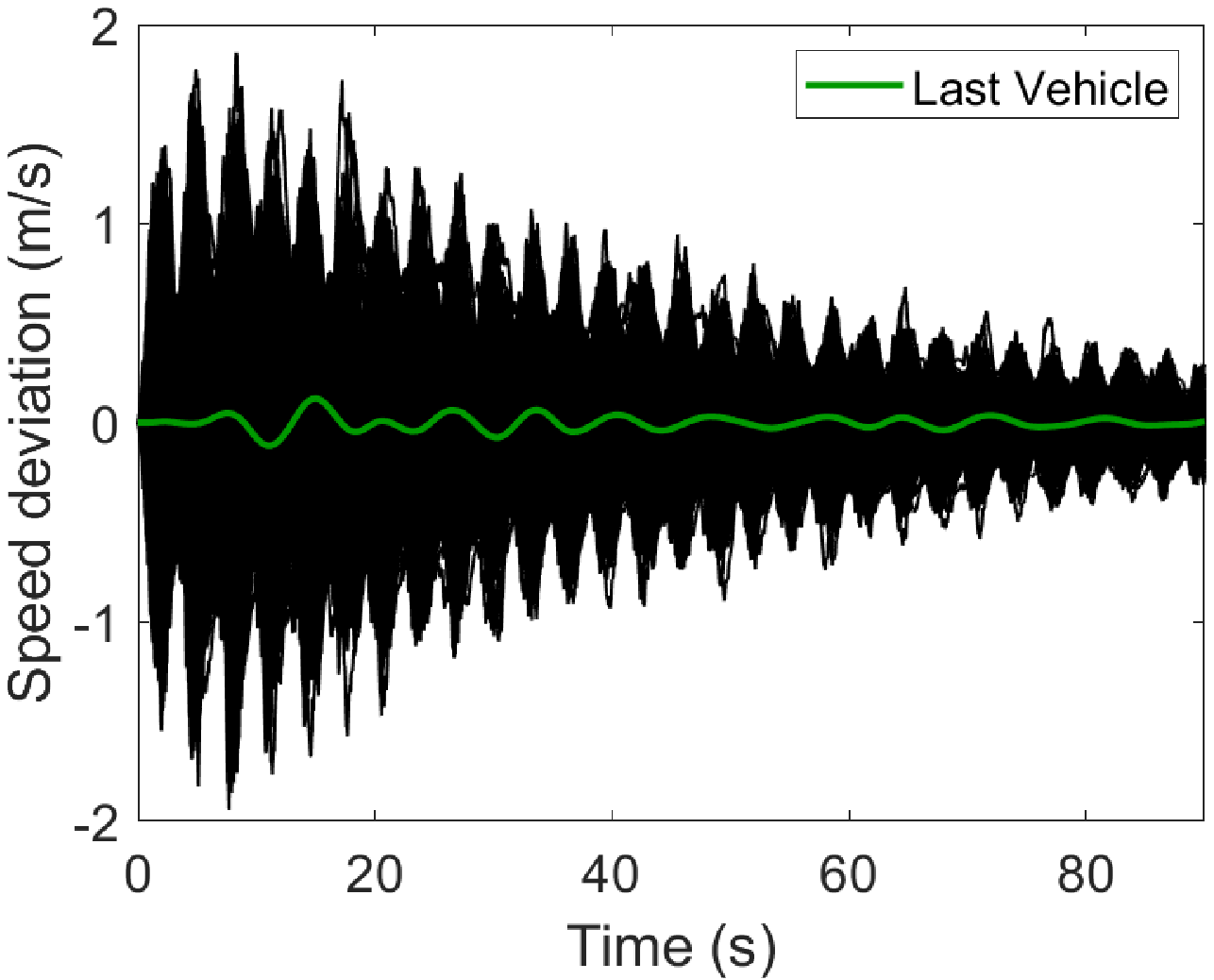}} \\
{\includegraphics[width=0.23\textwidth]{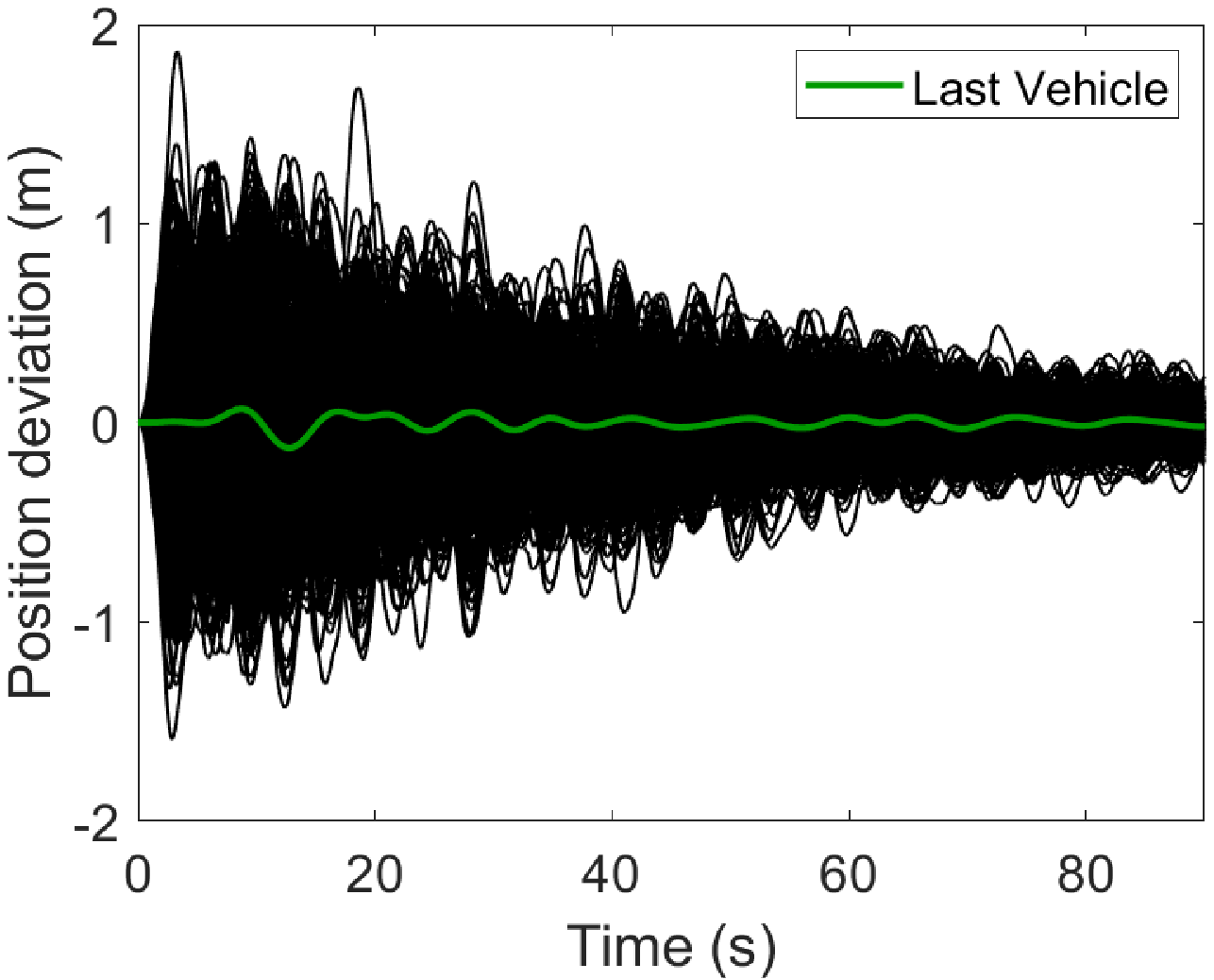}}&
{\includegraphics[width=0.23\textwidth]{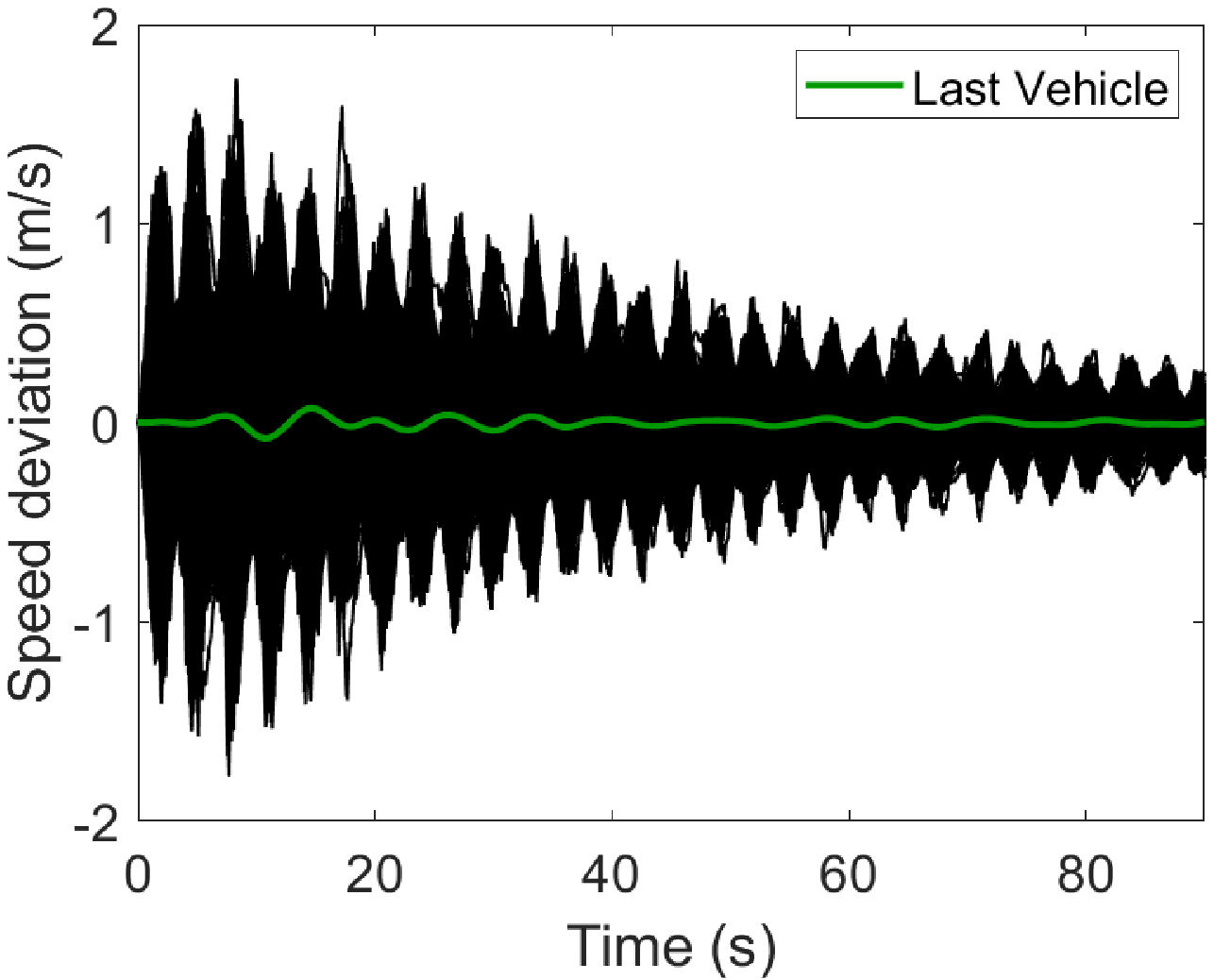}} 
\end{tabular}
\caption{\textcolor{black}{Simulations for the platoon system of Section \ref{sec:3}. Top panel: position/speed perturbations when $\varepsilon_i = 0$, $\forall i$. Bottom panel: speed/position perturbations with $\varepsilon_i = 1$, $\forall i$.}}\label{fig_1}
\end{figure} 
}

\section{Conclusions}\label{sec:conclusions}
We presented a sufficient condition for the $\mathcal{L}_{\infty}$ string stability of asymmetrically coupled bidirectional heterogeneous, nonlinear, platoon systems. Our result directly links  string stability to the design of the coupling protocols. We showed, via an example, how our result can be recast as an optimization problem and how this formulation can be used to design distributed control protocols for $\mathcal{L}_{\infty}$ string stable platoon systems. Future work will be aimed at studying: (i) whether automated vehicles can coexist with manually-driven vehicles and designing distributed control protocols supporting this mixed scenario; \textcolor{black}{(ii) the possibility of devising a fully distributed control protocol for platoon systems by e.g. making use of feed-forward terms and/or nonlinear spacing policies.}


\appendix
\section*{Appendix}
\section{Mathematical tools}\label{math_tools}

\noindent Let $A$ be a complex $n\times n$ matrix. We recall that the matrix measure of the matrix $A$ induced by a $p$-vector norm, $\abs{\cdot}_p$, is defined as $\mu_p(A) := \lim_{h \rightarrow 0^+} \frac{1}{h} \left(\norm{I+hA}-1\right)$, see \citep{Vid_93} \bb{and \citep{russo2010global} where matrix measures are used in the context of nonlinear contraction analysis}. In this paper we state our results in terms of $\mu_2(A) := \max_{i} \left( \lambda_i\left\{\frac{A+A^T}{2}\right\}\right)$, i.e. the matrix measure induced by the $2$-vector norm. Recall here that $p$-vector norms are monotone, i.e. $\forall x_1, x_2 \in\R^n$ such that $0 \le x_1 \le x_2$ it happens that $\abs{x_1}_p \le \abs{x_2}_p$ (where $x_1 \le x_2$ is understood component-wise). We make use of the following result from~\citep{Rus_diB_Son_13}.
\begin{Lemma}\label{lemma:norm}
Let: (i) $\abs{\cdot}_S$ and $\mu_S(\cdot)$ be, respectively, any $p$-vector and its induced matrix measure on $\R^N$; (ii) $\abs{\cdot}_G$ be the vector norm on $\R^{nN}$ defined as $\abs{\xi}_G := \abs{[\abs{\xi_1}_{L_1},\ldots,\abs{\xi_N}_{L_N}]}_S$; (iii) $\mu_G(\cdot)$ be the matrix measure induced by $\abs{\cdot}_G$. Finally, let $A:= (A_{ij})_{i,j = 1}^N \in\R^{nN\times nN}$, with $A_{ij} \in \R^{n\times n}$ and let $\hat A:= (\hat A_{ij})_{i,j = 1}^N \in\R^{N\times N}$, with $\hat A_{ii} := \mu_{L_i}(A_{ii})$ and $\hat A_{ij} := \norm{A_{ij}}_{L_{i,j}}$, $\forall i,j = 1,\ldots,N$. Then, $\mu_G(A) \le \mu_S(\hat A)$.
\end{Lemma}

\section{Proofs of the technical results}\label{sec:proof}

\subsection*{Proof of Theorem \ref{thm:predecessor_follower_general}}
For the sake of convenience we rewrite (\ref{eqn:unperturbedext}) in a more compact form as $\dot y_i = g_i(t,Y)$, $i =1,\ldots,N$, with 
\begin{equation}\label{eqn:gis}
\begin{split}
g_i(t,Y) & := f_i(y_i) + \tilde h_{i,i-1}(t,y_{i-1},y_i) \\
& + \varepsilon_i \tilde h_{i,i+1}(t,y_{i+1},y_i) + \tilde h_{i,0}(t,y_{i},x_0).
\end{split}
\end{equation}
Also, we rewrite \eqref{eqn:generalext} - \eqref{eqn:automated_asymmetric_general} as $\dot x_i =  g_i(t,X) + d_i(t)$, $i = 1,\ldots,N$, with the functions $g_i$'s defined as in (\ref{eqn:gis}).

Condition {\bf C1} implies that $Y^{\ast}(t)$ is a solution of the unperturbed dynamics (\ref{eqn:unperturbedext}). Let $z_i(t) := x_i(t) - y_i^{\ast}(t)$, $Z(t) := [z_1(t)^T,\ldots,z_N(t)^T]^T$ and $d(t)$ being the stack of all $d_i$'s (if the disturbance does not affect the $i$-th vehicle, then $d_i(t) = 0$). \textcolor{black}{Now, following Theorem A in \citep{1083507} (see also Theorem 3 in \citep{7403435} for a self-contained proof)}, the dynamics of $Z(t)$ can be expressed as
\begin{equation}\label{eqn:z_dynamics}
\dot Z(t) = A(t) Z(t) + d(t),
\end{equation}
with $A(t) := \int_0^1\tilde J(t,\eta X + (1-\eta)Y)d\eta$ and where \textcolor{black}{$\tilde J(t,X) :=\frac{\partial G}{\partial X}$}, with $G(t,X) := [g_1^T(t,X),\ldots,g_N^T(t,X)]^T$. \textcolor{black}{Then, as shown in \citep{1083507} and \citep{7403435}, one gets}
\begin{equation}\label{Dini2}
\begin{split}
	D^{+}\abs{Z(t)} = &  \mu(A(t))\abs{Z(t)} +\abs{d(t)}, 
\end{split}
\end{equation}
where  $D^{+}{\abs{Z(t)}}$ is the Dini derivative of $\abs{Z(t)}$, i.e.  $D^{+}{\abs{Z(t)}}:= \lim_{h\rightarrow 0^+}\sup \frac{1}{h}\left(\abs{Z(t+h)} - \abs{Z(t)}\right)$. Inequality (\ref{Dini2}) is valid for any vector norm and, in particular, it also holds when $\abs{Z} = \abs{Z}_G :=  \abs{\left[\abs{z_1}_{\textcolor{black}{2}},\ldots,\abs{z_N}_{\textcolor{black}{2}}\right]}_{\textcolor{black}{\infty}}$. That is, by definition, \textcolor{black}{$\abs{Z}_G = \sup_i\abs{z_i}_2$.}
Now, the rest of the proof is aimed at showing that there exists some $\bar c \ne 0$ such that $\mu_G\left(\tilde J(t,X)\right) \le - \bar c^2$, $\forall t \ge t_0$ and $\forall X$ (indeed, by means of subadditivity of matrix measures this implies that $\mu(A(t)) \le -\bar c^2$). In order to show this, partition the matrix $\tilde J$ in $\tilde J  = (\tilde J_{ij})_{i,j=1}^N$, with $\tilde J_{ij} \in \R^{n \times n}$. Then, by means of Lemma \ref{lemma:norm}, we have that $\mu_G( \tilde J) \le \mu_{\textcolor{black}{\infty}}(\hat J)$, where $\hat J  := (\hat J_{ij})_{i,j=1}^N \in \R^{N \times N}$:
\begin{equation}\label{eqn:blocks_def}
\begin{array}{*{20}l}
\hat J_{ii} = \mu_{\textcolor{black}{2}}\left(\tilde J_{ii}\right), & i =1,\ldots,N,\\
\hat J_{i,i+1} = \varepsilon_i \norm{\tilde J_{i,i+1}}_{\textcolor{black}{2}}, & i =1,\ldots,N-1,\\
\hat J_{i+1,i} = \norm{\tilde J_{i+1,i}}_{\textcolor{black}{2}}, & i =1,\ldots,N-1.
\end{array}
\end{equation}
For convenience, in (\ref{eqn:blocks_def}) and in what follows we are omitting the dependencies of the matrices $\tilde J_{ij}$'s on the state variables. Now, in order to show the result we need to show that there exists some $\bar c \ne 0$ such that $\mu_{\textcolor{black}{\infty}}(\hat J) \le -\bar c^2$, $\forall t \ge t_0$ and $\forall X$. \textcolor{black}{Now, by definition on $\mu_{\infty}(\cdot)$,  this is a row-dominance condition on the matrix $\hat J$.} That is, we need to show that there exists some $\bar c \ne 0$ such that, $\forall X$:
\textcolor{black}{\begin{equation}\label{eqn:ineq}
\begin{array}{*{20}l}
\mu_{2}\left(\tilde J_{ii}\right) + \varepsilon_i \norm{J_{i,i+1}}_2 + \norm{J_{i,i-1}}_2 \le -\bar c^2 , & i =1,\ldots,N,\\
\end{array}
\end{equation}
where we used the definition of the matrix $\hat J$} and, in order to make the notation more compact, we set $\norm{\tilde J_{i,j}}_{\textcolor{black}{2}} = 0$ whenever $i,j \notin\{1,\dots, N\}$. Now, by means of {\bf C2}, \textcolor{black}{the above expression can be upper bounded, for all $i=1,\ldots,N$, by}: $-c^2 + \left(1 + \max_i\varepsilon_i\right) \bar J:=- \bar{c}^2$.
In turn, from {\bf C3} we get  $-c^2 + \left(1 + \max_i\varepsilon_i\right) \bar J < 0$, thus implying that there exists some $\bar c \ne 0$ such that $\mu_{\textcolor{black}{\infty}}(\hat J) \le -\bar c^2$, $\forall X$. Together with (\ref{Dini2}), this implies that:
\begin{equation}\label{Dini3}
\begin{split}
	D^{+}{\abs{Z(t)}}_{\textcolor{black}{G}}&\le -\bar c^2\abs{Z(t)}_{\textcolor{black}{G}} +\abs{d(t)}_{\textcolor{black}{G}}.
\end{split}
\end{equation}
\textcolor{black}{From (\ref{Dini3}) we get
$D^{+}{\abs{Z(t)}}_{\textcolor{black}{G}} \le -\bar c^2\abs{Z(t)}_{\textcolor{black}{G}} +\textcolor{black}{\sup_i\norm{d_i(\cdot)}_{\mathcal{L}_\infty}}$, where we used the definition of $\abs{\cdot}_G$ together with the definition of the supremum norm.} Thus, application of the Gronwall's inequality yields: 
$$
\abs{Z(t)}_{\textcolor{black}{G}} \le e^{-\bar c^2(t-t_0)}\abs{Z(t_0)}_{\textcolor{black}{G}} + \frac{1- e^{-\bar c^2(t-t_0)}}{\bar c^2}\textcolor{black}{\sup_i\norm{d_i(\cdot)}_{\mathcal{L}_\infty}},
$$ 
$\forall t\ge t_0$. Finally, since $Z(t) := X(t) - Y^{\ast}(t)$, this yields, \textcolor{black}{by the definition of $\abs{\cdot}_G$}, 
\textcolor{black}{\begin{equation*}
\begin{split}
 \sup_i\abs{x_i(t) -y_i^{\ast}(t)}_{\textcolor{black}{2}} & \le e^{-\bar c^2(t-t_0)}\sup_i\abs{x_i(t_0)- y_i^{\ast}(t_0)}_{\textcolor{black}{2}}\\
 & + \frac{1- e^{-\bar c^2(t-t_0)}}{\bar c^2}\textcolor{black}{\sup_i\norm{d_i(\cdot)}_{\mathcal{L}_\infty}},
\end{split}
\end{equation*}}
$\forall t\ge t_0$ and this  gives the result.\qed
 
\subsection*{Proof of Corollary \ref{thm:string_automatic_vehicles}}

Apply, to the dynamics (\ref{eqn:general}) - (\ref{eqn:automated}), the coordinate transformation $\tilde x_i := T_ix_i$, where $T_i$ is given as in (\ref{eqn:T_matrix}). This yields the transformed dynamics:
\begin{equation}\label{eqn:transformed_dynamics}
\begin{split}
& \dot{\tilde{x}}_i = T_iFT_i^{-1}\tilde x_i  + T_iB \left(h_{i,i-1}(T_{i-1}^{-1}\tilde x_{i-1},T_{i}^{-1}\tilde x_{i}) + \right.\\
&\left.+\varepsilon_ih_{i,i+1}(T_{i+1}^{-1}\tilde x_{i+1},T_{i}^{-1}\tilde x_{i})+ \textcolor{black}{h_i^{(0)}}(T_{i}^{-1}\tilde x_{i},x_0)\right) + \tilde d_i(t), 
\end{split}
\end{equation}
with $B := [0, 1]^T$ and $\tilde d_i(t) = T_i d_i(t)$. Let $\tilde{y}_{d,i}:= T_i y_{d,i}$, then the unperturbed dynamics of (\ref{eqn:transformed_dynamics}) is
\begin{equation}\label{eqn:transformed_dynamics_unperturbed}
\begin{split}
& \dot{\tilde{y}}_i = T_iFT_i^{-1}\tilde y_i  + T_iB \left(h_{i,i-1}(T_{i-1}^{-1}\tilde y_{i-1},T_{i}^{-1}\tilde y_{i}) + \right.\\
&\left.+\varepsilon_ih_{i,i+1}(T_{i+1}^{-1}\tilde y_{i+1},T_{i}^{-1}\tilde y_{i})+ \textcolor{black}{h_i^{(0)}}(T_{i}^{-1}\tilde y_{i},x_0)\right). 
\end{split}
\end{equation}
Now, it suffices to note that: (i) the fulfillment of {\bf C1} of Corollary \ref{thm:string_automatic_vehicles} implies the fulfillment of {\bf C1} of Theorem \ref{thm:predecessor_follower_general}; (ii) differentiation of (\ref{eqn:transformed_dynamics}) yields the Jacobian matrix  $J  := (J_{ij})_{i,j=1}^N \in \R^{2N \times 2N}$, where each element $J_{ij}\in\R^{2\times 2}$ is given by (\ref{eqn:big_Jacobian}). In turn, this means that the fulfillment of conditions {\bf C2} - {\bf C3} of Corollary \ref{thm:string_automatic_vehicles} implies that the same conditions of Theorem \ref{thm:predecessor_follower_general} are also fulfilled for the dynamics (\ref{eqn:transformed_dynamics}).

Thus, application of Theorem \ref{thm:predecessor_follower_general} to the dynamics (\ref{eqn:transformed_dynamics}) with $y_i^{\ast}(t) = \tilde{y}_{d,i}(t)$ yields \textcolor{black}{$\sup_i\abs{\tilde x_i(t) -\tilde y_{d,i}(t)}_{\textcolor{black}{2}} \le e^{-\bar c^2t}\sup_i\abs{\tilde x_i(0)- \tilde y_{d,i}(0)}_{\textcolor{black}{2}} + \frac{1- e^{-\bar c^2t}}{\bar c^2}\sup_i\norm{\tilde d_i(\cdot)}_{\mathcal{L}_\infty}$}, $\forall t \ge 0$. Now:
\textcolor{black}{\begin{equation}\label{eqn:part_1}
\begin{split}
 \sup_i\abs{\tilde x_i(t) -\tilde y_{d,i}(t)}_{\textcolor{black}{2}} \ge \underline{\lambda}  \sup_i\abs{x_i(t) -y_{d,i}(t)}_{\textcolor{black}{2}},
\end{split}
\end{equation}}
where $\underline{\lambda} := \min_i\left\{\sigma_{\min}(T_i)\right\}$. Also:
\textcolor{black}{\begin{equation}\label{eqn:part_2}
\begin{split}
\sup_i\abs{\tilde x_i(0)- \tilde y_{d,i}(0)}_{\textcolor{black}{2}} & \le\bar{\lambda}\sup_i\abs{ x_i(0)- y_{d,i}(0)}_{\textcolor{black}{2}},
\end{split}
\end{equation}}
 where $\bar{\lambda} := \max_i\left\{\sigma_{\max}(T_i)\right\}$. Finally, we have:
\textcolor{black}{\begin{equation}\label{eqn:part_3}
\begin{split}
\sup_i\norm{\tilde d_i(\cdot)}_{\mathcal{L}_\infty} \le \max_i\left\{\sigma_{\max}(T_i)\right\}\sup_i\norm{d_i(\cdot)}_{\mathcal{L}_\infty}.
\end{split}
\end{equation}}
Then, (\ref{eqn:upper_bound}) directly follows from (\ref{eqn:part_1}) - (\ref{eqn:part_3}).  \qed

\section{Recasting {\bf C2} and {\bf C3} as an optimization problem}\label{sec:opt}
Formally, finding the set of control gains $K_{vi}$, $K_{vi,0}$, $K_{pi,0}$, $K_{pi,1}$, $K_{pi,2}$, fulfilling {\bf C2} and {\bf C3} can be recast as the following optimization problem:
\begin{equation}\label{eqn:opt_general}
\begin{split}
&\min_{\begin{array}{*{20}c}
K_{vi},K_{pi,0},K_{pi,1},\\
K_{pi,2},K_{vi,0},\alpha_i,C,\bar J,\varepsilon_i\end{array}} \mathcal{J}\left(K_{vi}, K_{vi,0}, K_{pi,0}, K_{pi,1}, K_{pi,2},\varepsilon_i\right)\\
&\text{s.t.} \\
& K_{vi}>0,\quad K_{pi,0}>0, \quad K_{vi,0}>0, \quad K_{pi,1}>0, \\
& K_{pi,2}>0, \quad \bar J >0, \quad C >0, \quad \alpha_i >0, \quad 0 \le \varepsilon_i \le 1\\
&\mu_{2}\left(J_{i,i}(\alpha_i,x_i,x_{i-1},\varepsilon_i)\right) \le -C, \quad \left(\varepsilon_i+1\right)\bar J<C,\\
&\norm{J_{i,i-1}(\alpha_i,x_i,x_{i-1})}_{\textcolor{black}{2}}\le \bar J,\quad \norm{J_{i,i+1}(\alpha_i,x_i,x_{i+1})}_{\textcolor{black}{2}} \le \bar J,\\
\end{split}
\end{equation}
where the decision variables are the control gains and the auxiliary variables $\alpha_i$, $C$, $\bar J$. We set $\mathcal{J}\left(\cdot\right) = -\bar{g_i} = -K_{pi,1}K_{pi,2}$ and solve the above problem for fixed $\alpha_i >0$ and $0\le\varepsilon_i\le 1$. \bb{With this choice of the cost function, the upper bound of $\partial g_i / \partial x$ is maximized (note that other cost functions can be considered as the steps described below are not dependent on $\mathcal{J}(\cdot)$). Now, we recast the constraints in (\ref{eqn:opt_general}) as LMIs, see e.g.~\citep{doi:10.1137/1.9781611970777}: (i) by definition, the constraint $\mu_{2}\left(J_{i,i}\right)\leq -C$ is equivalent to $[J_{i,i}]_s \prec -CI_2$; (ii) by definition, the constraint $\norm{J_{i,i-1}(\alpha_i,x_i,x_{i-1})}_{\textcolor{black}{2}}\le \bar J$ is equivalent to $\bar J^2 I_2- J_{i,i-1} J_{i,i-1}^{T}\succeq 0$ and hence, by means of the Schur complement, see e.g.~\citep[Theorem $7.7.7$]{Hor_Joh_99} and dividing by $\bar J > 0$, this is in turn equivalent to $\begin{bmatrix}\bar J \cdot I_2&J_{i,i-1}\\  J_{i,i-1}^{T} & \bar J\cdot I_2\end{bmatrix}\succeq 0$. Moreover, as $J_{i,i-1}$ and $J_{i,i}$ both depend linearly on $\partial g_i / \partial x$ and $0\leq \partial g_i / \partial x\leq \bar{g_i}$, then the above constraints define convex sets. Therefore: (i) $[J_{i,i}]_s \prec -CI_2$ can be replaced by the pair of constraints $[J_{i,i,lb}]_s\prec -CI_2$ and $[J_{i,i,ub}]_s\prec -CI_2$; (ii) $\norm{J_{i,i-1}}_{2}\leq \bar J$ can be replaced by the pair of constraints $\norm{J_{i,i-1,lb}}_2\leq \bar J$ and $\norm{J_{i,i-1,ub}}_2\leq \bar J$ (see (\ref{eqn:constraints_matrix}) below for the definition of the matrices). This yields to the convex optimization problem solved in Section \ref{sec:3}:
\begin{equation}\label{optimisation}
\begin{split}
&\min_{K_{vi},K_{pi,0},K_{vi,0},\bar{g_i},C,\bar J} -\bar{g_i}, \quad
\text{subject to:} \\
&\quad K_{vi}>0,\quad K_{pi,0}>0, \quad K_{vi,0}>0, \quad \bar{g_i}>0, \quad \bar J >0, \\
&\quad C >0, \quad \alpha_i >0, \quad 0 \le \varepsilon_i \le 1, \quad \left(\varepsilon_i+1\right)\bar J-C < 0,  \\
& \begin{bmatrix}\bar J \cdot I_2& J_{i,i-1,lb}\\ J_{i,i-1,lb}^{T} & \bar J\cdot I_2\end{bmatrix}\succeq 0, \quad
\begin{bmatrix}\bar J \cdot I_2& J_{i,i-1,ub}\\ J_{i,i-1,ub}^{T} & \bar J\cdot I_2\end{bmatrix}\succeq 0,\\
&\quad[J_{i,i,lb}]_s\prec - C I_2, \quad[J_{i,i,ub}]_s\prec - C I_2.
\end{split}
\end{equation}
In our implementation in Section \ref{sec:3}, the above problem was solved numerically for different values of $\alpha_i$ and $\varepsilon_i$. For any choice of such parameters, the solver was always able to converge to an optimal solution, thus returning a set of control gains minimizing the cost function. In the simulations of Section \ref{sec:3} we made use of the set of control gain that was returning the lowest value of the cost function across all the numerical experiments.} The files implementing the optimization problem can be made available upon request.  

\begin{figure*}[b]
\hrulefill
\begin{equation}\label{eqn:constraints_matrix}
\begin{split}
&[J_{i,i,lb}]_s:=\begin{bmatrix}-\alpha_i K_{pi,0}&1+\alpha_i^2K_{pi,0} - \alpha_i\left(\left(1+\varepsilon_i\right)K_{vi} + K_{vi,0}\right)\\
-K_{pi,0} & \alpha_iK_{pi,0} - (1+\varepsilon_i)K_{vi}- K_{vi,0}\end{bmatrix}, \quad \quad \quad \quad \quad \quad \quad \quad J_{i,i-1,lb} := \begin{bmatrix}
0 & \alpha_i K_{vi}\\
0 & K_{vi}
\end{bmatrix}\\
 &[J_{i,i,ub}]_s:=
\begin{bmatrix}-\alpha_i K_{pi,0}&1+\alpha_i^2\left((1+\varepsilon_i)\bar{g_i}+K_{pi,0}\right) - \alpha_i\left(\left(1+\varepsilon_i\right)K_{vi} + K_{vi,0}\right)\\
-K_{pi,0} & \alpha_i \left((1+\varepsilon_i)\bar{g_i}+K_{pi,0}\right) - (1+\varepsilon_i)K_{vi}- K_{vi,0}\end{bmatrix},
\quad
J_{i,i-1,ub} := \begin{bmatrix}
\alpha_i\bar g_i &-\alpha_i^2\bar g_i +\alpha_i K_{vi}\\
\bar g_i & -\alpha_i\bar g_i + K_{vi}
\end{bmatrix}.\\
\end{split} 
\end{equation} 
\end{figure*}

\bibliographystyle{elsarticle-harv}
\bibliography{russobib2a}

\end{document}